\documentclass[preprintnumbers,aps,twocolumn,floatfix,superscriptaddress]{revtex4-1}
\usepackage{eso-pic,calc}
\usepackage{graphicx}
\usepackage{amsmath, amsthm, amssymb}
\usepackage{epsfig}
\usepackage{bm}
\usepackage{color}
\usepackage[colorlinks,linktocpage,linkcolor=black]{hyperref}

\def\beqr{\begin{eqnarray}}
\def\eqnr{\end{eqnarray}}
\def\beq{\begin{equation}}
\def\bc{\begin{center}}
\def\ec{\end{center}}
\def\eqn{\end{equation}}

\def\rmp#1#2#3{{ Rev. Mod. Phys.} {\bf #1}, #2 (#3)}
\def\prl#1#2#3{{ Phys. Rev. Lett.} {\bf #1}, #2 (#3)}

\def\pre#1#2#3{Phys. Rev. E {\bf #1}, #2 (#3)}

\def\prx#1#2#3{Phys. Rev. X {\bf #1}, #2 (#3)}
\def\pnas#1#2#3{Proc. Natl. Acad. Sci. (USA) {\bf #1}, #2 (#3)}

\def\natphys#1#2#3{Nat. Phys. {\bf #1}, #2 (#3)}

\def\etl{$et~al.$~}

\begin{document}

\title{Finite-size scaling of critical avalanches}

\author{Avinash Chand Yadav}
\affiliation{Department of Physics, Institute of Science, Banaras Hindu University, Varanasi 221005, India}

\author{Abdul Quadir}
\affiliation{Department of Physics, Aligarh Muslim University, Aligarh 202002, India}

\author{Haider Hasan Jafri}
\affiliation{Department of Physics, Aligarh Muslim University, Aligarh 202002, India}

\begin{abstract}
{We examine probability distribution for avalanche sizes observed in self-organized critical systems. While a power-law distribution with a cutoff because of finite system size is typical behavior, a systematic investigation reveals that it may decrease on increasing the system size at a fixed avalanche size. We implement the scaling method and identify scaling functions. The data collapse ensures a correct estimation of the critical exponents and distinguishes two exponents related to avalanche size and system size. Our simple analysis provides striking implications. While the exact value for avalanches size exponent remains elusive for the prototype sandpile on a square lattice, we suggest the exponent should be 1. The simulation results represent that the distribution shows a logarithmic system size dependence, consistent with the normalization condition. We also argue that for train or Oslo sandpile model with bulk drive, the avalanche size exponent is slightly less than 1 that is significantly different from the previous estimate 1.11.} 
\end{abstract}

\maketitle

\section{Introduction}
The emergent scale-invariant feature~\cite{Gros_2014, Miguel_2018, S_Thurner_2018, Palmieri_2020_1, Sethna_2001, Nagler_2020} remains one of the most remarkable observation occurring in systems as diverse as friction factor in turbulent flow~\cite{Goldenfeld_2006}, jamming transition~\cite{Sethna_2016}, and phylogenetic trees topology~\cite{Goldenfeld_2020}, to name a few. While the scaling behavior reflects a lack of a characteristic scale, diverse systems may have the same scaling exponents irrespective of their different microscopic dynamics. The notion of universality classes plays an important role, and the intriguing nature of the scaling feature continues to attract attention.

Such features can arise near the critical point of a continuous transition between order and disorder phases. The hypothesis of self-organized criticality (SOC)~\cite{Bak_1987, Bak_1988, Bak_1996, Pruessner_2012}, poised by Bak, Tang, and Wiesenfeld (BTW), explains the underlying origin of scaling in natural systems, which remain far away from equilibrium. 
According to SOC, a class of spatially extended driven-dissipative systems spontaneously organizes into a critical state. The response to a noisy drive exhibits non-linearity, and the fluctuations, termed \emph{critical avalanches}, show scaling in size probability distribution function (PDF). So far, SOC has explained scaling features in a broad range of phenomena spanning from earthquake~\cite{Olami_1992} to biological evolution~\cite{Bak_1993} and neuronal avalanches~\cite{Levina_2007, Kossio_2018,Zeraati_2021}.

Despite considerable efforts, many aspects of SOC remain yet not clearly understood.
One fundamental goal is to determine the scaling exponent for avalanches size distribution. In this context, the exact value of the exponent remains elusive for the paradigmatic BTW sandpile on the square lattice. Initial numerical estimates suggest $\tau \approx 1$~\cite{Bak_1987}. Subsequently, Zhang~\cite{Zhang_1989} proposed a scaling theory that correctly justified the observation of BTW. However, Manna later performed large system size simulations and found $\tau \sim 1.2$~\cite{Manna_1990}. Since then, several studies have examined this issue using different tools like the mean-field~\cite{Christensen_1993} and renormalization group methods~\cite{Zapperi_1994}. Decomposing an avalanche into a sequence of `waves' of toppling, Priezzhev \etl \cite{Priezzhev_1996} argued $\tau = 6/5$. However, the legitimacy of underlying assumptions has been questioned by Paczuski \etl \cite{Paczuski_1997}.

Typically, such systems are not analytically tractable, and the numerical estimation of the exponents comes from simulation studies~\cite{Usadel_1997}. It is desirable to determine the scaling exponents accurately to validate theoretical arguments and universality classes. The commonly accepted approach is to get data collapse from finite-size scaling (FSS), for which a systematic method is moment analysis~\cite{Lubeck_2000}. However, the FSS breaks to provide a good data collapse in various cases~\cite{Christensen_2005, Lubeck_2000_1, Lubeck_2007}. While the area distribution obeys FSS for BTW sandpile, the size distribution does not, particularly near cutoff. Examples also include bulk driven Oslo sandpile~\cite{Christensen_1996, Frette_1996, Paczuski_1996} or train model of the earthquakes~\cite{Burridge_1967, Vieira_2000}. The multi-fractal scaling and edge events appear to explain the breaking of FSS~\cite{Tebaldi_1999, Lise_2001}.

Our main contribution, in this paper, comprises the following. While the power-law distribution with a cutoff remains a well-recognized feature associated with critical avalanches, it may also be an explicit function of system size. We emphasize that the probability may decrease with the system size for a fixed avalanche size. Although such system size dependence is a simple feature, it is unclear both its systematic analysis and inferences. We show a simple scaling analysis that can capture this feature. The method relies on identifying the characteristics of scaling functions.

To observe the prominent feature, we numerically investigate several SOC models. We also point out implications and limitations encountered in numerical computations. The exact value of the exponent should be 1 for the BTW sandpile because the size distribution follows a logarithmic system size dependence. It is easy to follow from the normalization of the power-law PDF with an upper cutoff. We also note similar features in bulk driven train model.

The plan of the paper is as follows. In Sec.~\ref{sec_2}, we recall the BTW sandpile model and show simulation results for avalanche area and size distributions. Sec.~\ref{sec_3} presents an analysis revealing the system size  scaling. We show similar results for bulk driven train model in Sec.~\ref{sec_4}. Finally, Sec.~\ref{sec_5} provides a summary and discussion.

\section{BTW sandpile model} \label{sec_2}
To test the proposed scaling behavior, we examine several models manifesting SOC. These models explain avalanches observed in diverse systems, ranging from neuronal networks to earthquakes and sandpiles. Here, we first show results for the BTW sandpile model. 
Consider a square lattice with $N = L^2$ sites, where $L$ is the linear extent. Associate a discrete height or slope variable to each site as $z_i$ such that $0\le z_i < z_0$, where $z_0 = 4$ is the threshold. The system is driven by randomly selecting a site and updating it as $z_i \to z_i +1$. If a site $i$ is unstable $z_i\ge z_0$, the site relaxes as:   
\begin{eqnarray}
z_i \to z_i-4, \nonumber \\
z_j \to z_j+1, \nonumber
\end{eqnarray}
where $j$ denotes the nearest neighbors. As a result, the neighbor site(s) may become unstable. The relaxation continues until all sites become stable, and the open boundary allows dissipation. A new driving occurs when the ongoing avalanche is over. Thus, the timescale separation between drive and dissipation excludes interaction among avalanches.

\begin{figure}[t]
	\centering
	\scalebox{0.68}{\includegraphics{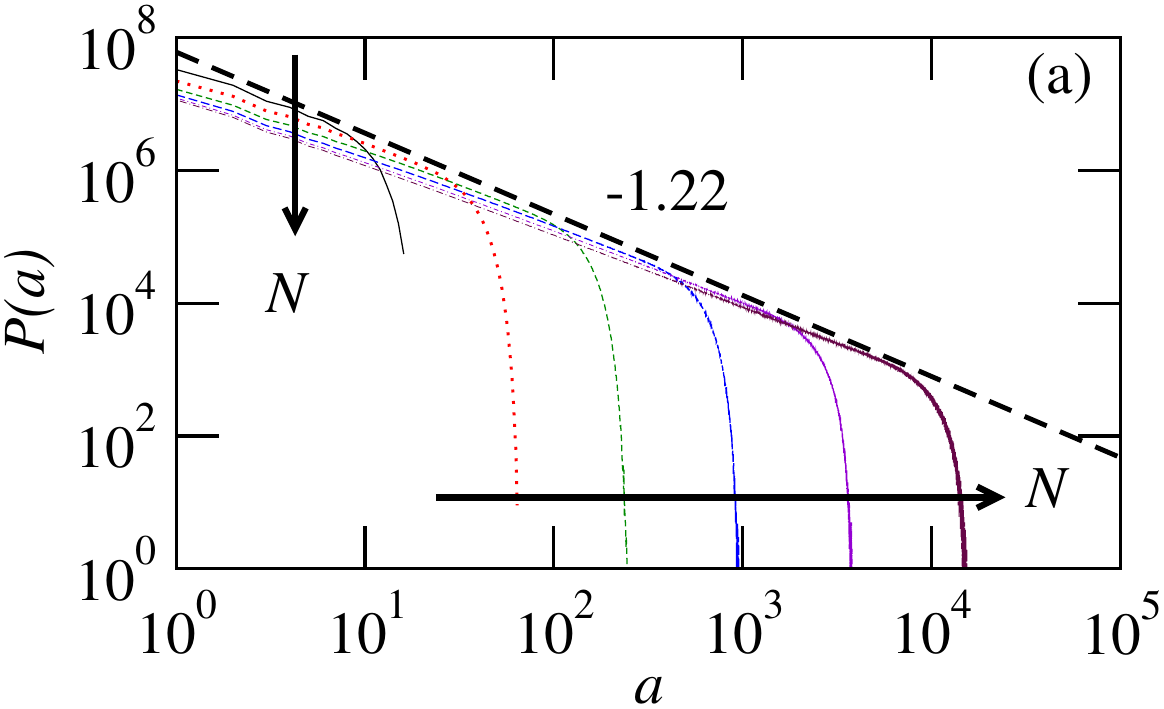}}
	\scalebox{0.69}{\includegraphics{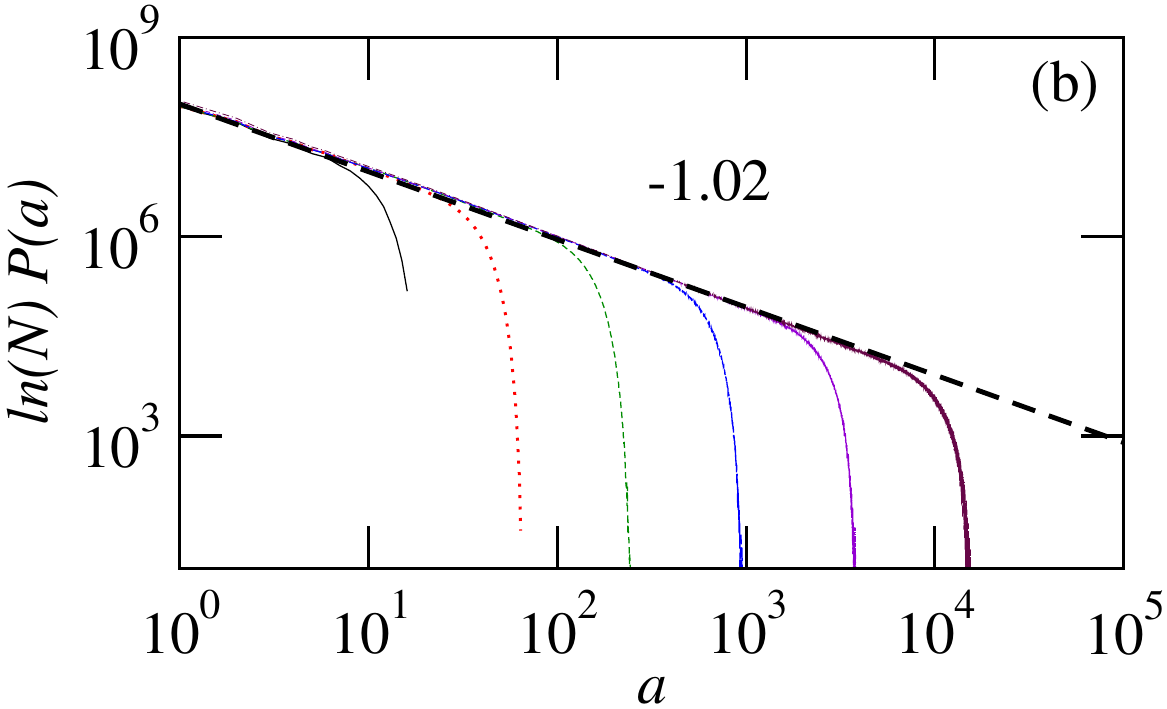}}
	\scalebox{0.65}{\includegraphics{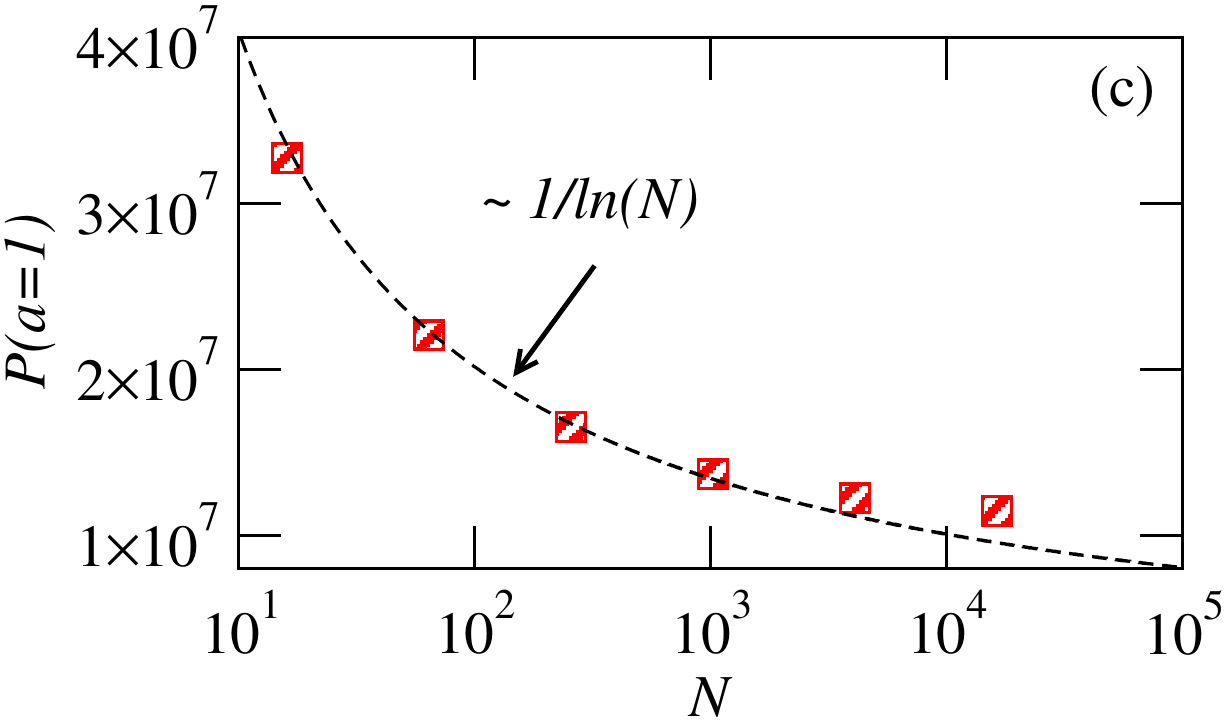}}	
	\caption{The curves show area distribution for BTW sandpile model, where (a) $P(a)$  and (b) $\ln(N) P(a)$. The system size $N$ varies from $2^4, 2^6, \cdots$ to $2^{14}$. Throughout our simulation, we use $\mathcal{M}=10^8$.   For comparison, we draw lines (thick dashed) with respective slopes. (c) The variation of $P(a=1)$ with $N$ along with best fit curve $\sim 1/\ln(N)$.  }
	\label{fig_2d_btw_pdf_area} 
\end{figure}

\begin{figure}[t]
	\centering
	\scalebox{0.68}{\includegraphics{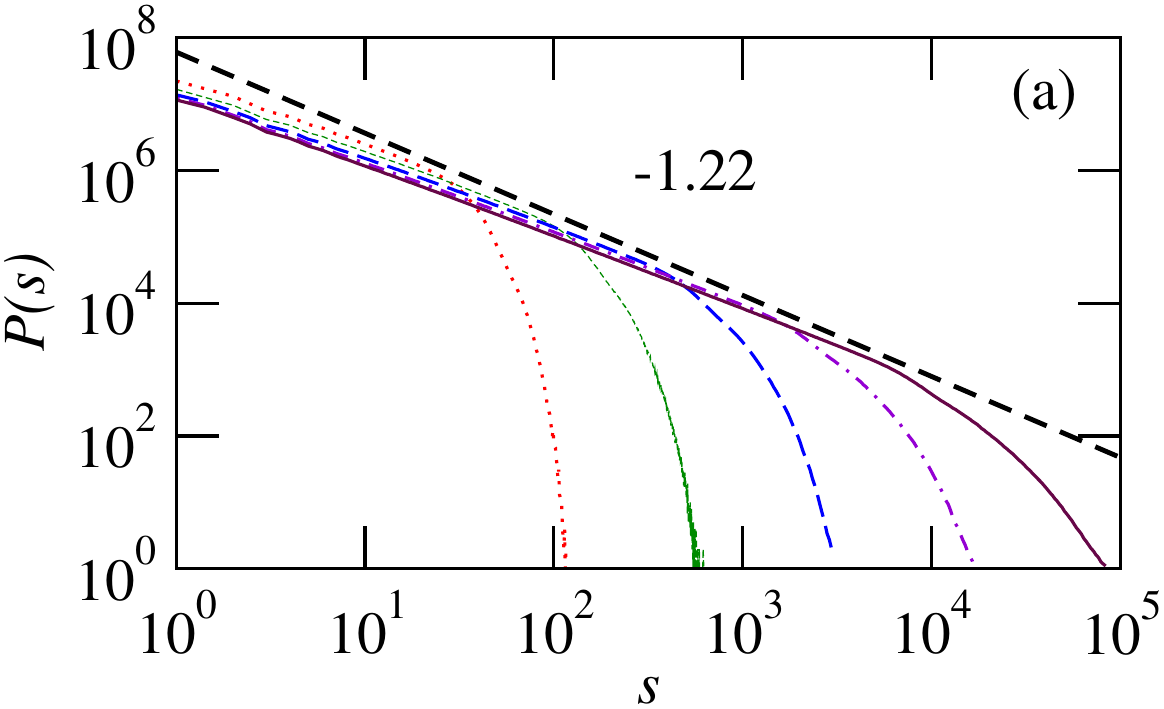}}
	\scalebox{0.69}{\includegraphics{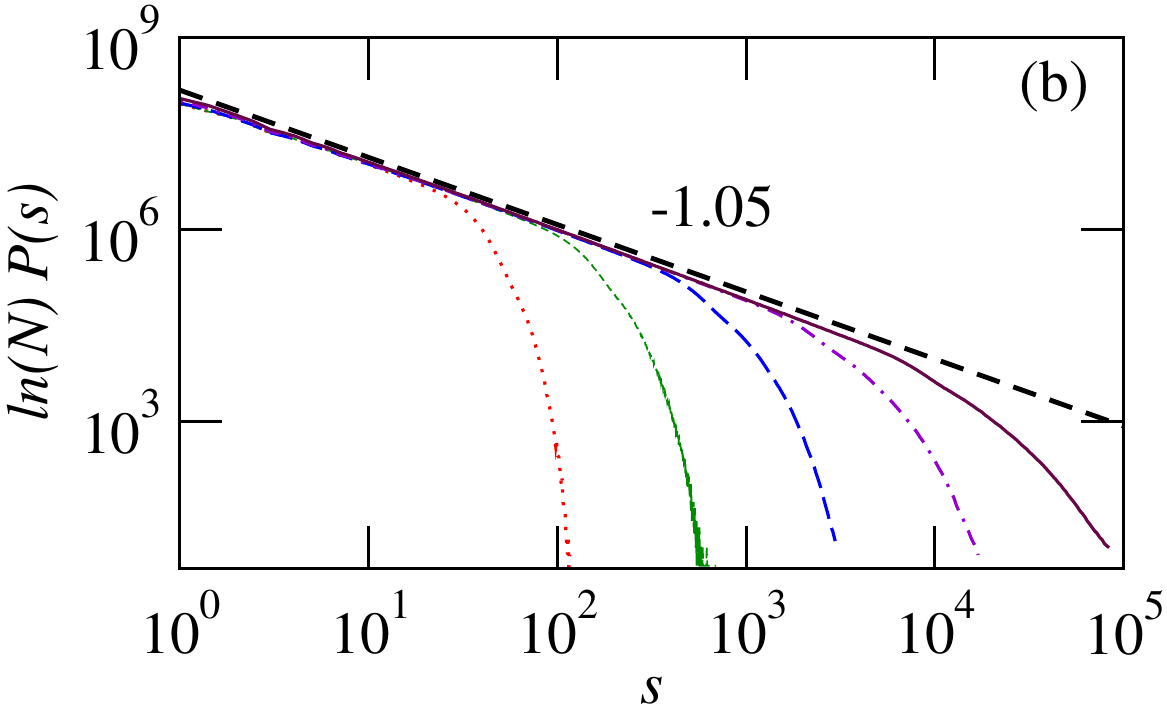}}	
	\caption{For BTW sandpile model, the plot of (a) $P(s)$ and (b) $\ln(N)P(s)$, with different system size $N$, varying from $ 2^6, 2^8 \cdots$ to $2^{14}$.}
	\label{fig_2d_btw_size_1} 
\end{figure}

The number of total (distinct) toppled sites denotes the avalanche size (area). We numerically examine the PDF for area $a$ and size $s$ variables. We can clearly see a system size effect for the area distribution in Fig.~\ref{fig_2d_btw_pdf_area} (a), and this vanishes if we plot $\ln(N) P(a)$ [cf. Fig.~\ref{fig_2d_btw_pdf_area} (b)]. Plotting $P(a=1)$ for different $N$, we note $\sim 1/\ln(N)$ type behavior [cf. Fig.~\ref{fig_2d_btw_pdf_area} (c)]. The mean area (not shown) scales as $\langle a\rangle \sim N/\ln(N)$. Similarly, Fig.~\ref{fig_2d_btw_size_1} supports the logarithmic feature for the avalanche size distribution.

\section{System size scaling}\label{sec_3}
Consider sandpile systems showing critical avalanches. Observable $x$ can describe the events like size (total toppled sites) $s$ and area (spatial extent of size) $a$. A systematic numerical investigation suggests that the probability distribution of the event $x$ obeys a decaying power-law behavior
\begin{equation} 
P(x, x_c) \sim \begin{cases} x_c^{-\theta}x^{-\tau_x}, ~~~~~~~~~~~~{\rm if}~ x\ll x_c,\\ {\rm rapid ~ decay},~~~~~~~~ {\rm if }~~ x\approx x_c, \end{cases} 
\label{eq_prob_1}
\end{equation}
where $x_c \sim L^{D_x}$, with $L$ being the linear extent of system and $D_x$ is cutoff exponent. Strikingly, Eq.~(\ref{eq_prob_1}) captures an unusual feature: The probability decreases on increasing the system size while keeping $x$ fixed [for example, cf. Fig.~\ref{fig_2d_btw_pdf_area} (a)]. We include a multiplicative pre-factor (a function of $x_c$) to account for the finite size effect. We assume one of the simplest forms $x_{c}^{-\theta}$, where $\theta$ is a scaling exponent.

Under what mathematical condition can we expect such  behavior, associated with power-law distribution? A well-defined probability density function must satisfy two conditions: (i) Positivity $0\le P(x)\le 1$ and (ii) normalization $\int_x P(x)dx = 1$. For simplicity, we consider a power-law PDF with a sharp cutoff $P(x) = A x^{-\tau_x}$, with $1\le x\le x_c$, where $A$ is a normalization factor. The normalization yields
\begin{equation}
A = \begin{cases}(1-\tau_x)/ \left[ x_{c}^{1-\tau_x}-1\right],~~~~~~{\rm if}~\tau_x\neq 1,\\ 1/\ln (x_c),~~~~~~~~~~~~~~~~~~~~~~~{\rm if}~\tau_x=1.\end{cases}\nonumber
\end{equation} 
For large but finite system, $x_c\gg 1$. In turn, the factor $A(x_c)$ can be expressed as  
\begin{equation}
A(x_c) \sim \begin{cases}1 + \mathcal{O}(x_{c}^{1-\tau_x}),~~~~~~~{\rm if}~\tau_x> 1,\\ x_{c}^{-\epsilon},~~~~~~~~~~~~~~~~~~~~{\rm if}~\tau_x=1, \\ 1/x_{c}^{1-\tau_x}, ~~~~~~~~~~~~~~{\rm if}~\tau_x<1.\end{cases}
\label{A_xc}
\end{equation}
Here, we use an approximation $\ln x_c \sim x_{c}^{\epsilon}$. Comparing Eqs.~(\ref{eq_prob_1}) and~(\ref{A_xc}), we can easily note the scaling exponent 
\begin{equation}
\theta = \begin{cases}0,~~~~~~~~~~~~{\rm if}~\tau_x> 1,\\ \epsilon,~~~~~~~~~~~~{\rm if}~~\tau_x=1, \\ 1-\tau_x, ~~~~~~{\rm if}~\tau_x<1.\end{cases}
\end{equation}

In the thermodynamic limit $x_c\to \infty$, the avalanches should show decaying power-law distribution $x^{-\tau_x}$ with $\theta = 0$ (a sign of generic criticality) and $\tau_x>1$. However, there is an upper cutoff for finite but large systems. As a result, the critical exponent $\tau_x$ can accept a value less than or equal to 1. If $\tau_x = 1$, a logarithmic behavior appears. This case is usually challenging to verify numerically since such an effect becomes too small to see for a large system size. If $\tau_x < 1$, then $\tau_x+\theta = 1$. Eventually,
\begin{equation}
\tau_x +\theta \geq 1.
\label{ineq_1}
\end{equation}

\begin{figure}[t]
	\centering
	\scalebox{0.69}{\includegraphics{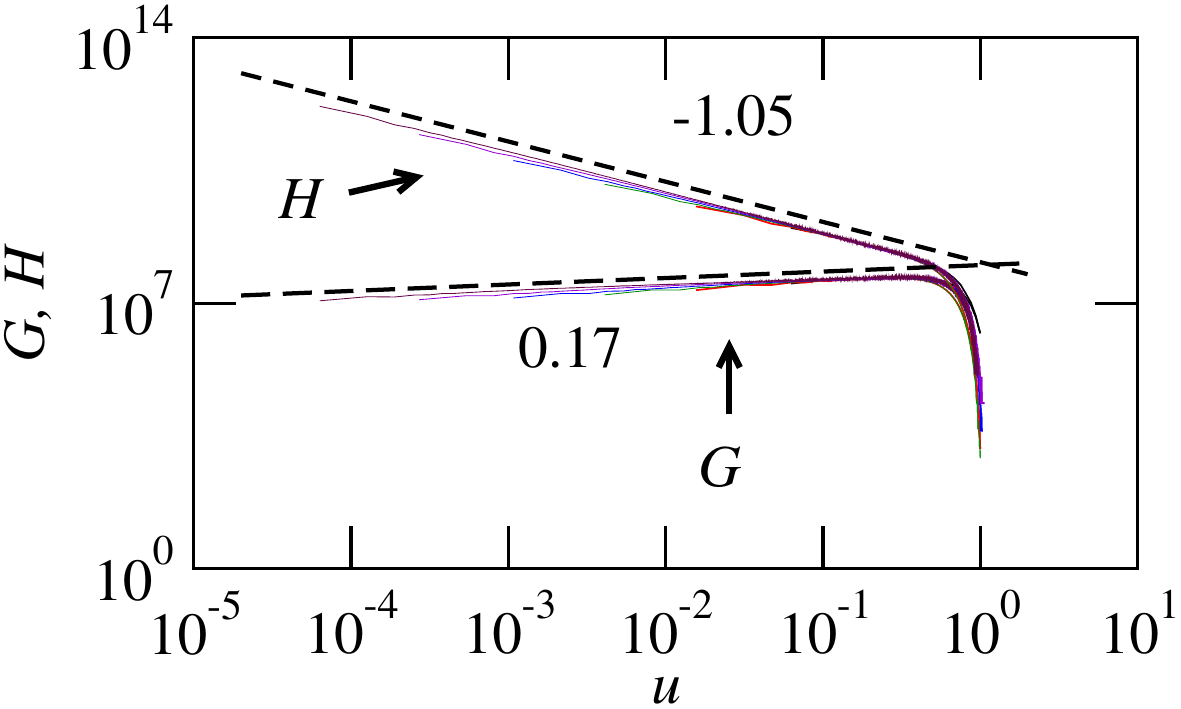}}
	\caption{The data collapses correspond to Fig.~\ref{fig_2d_btw_pdf_area} (a). The scaling exponent is $\theta \approx 0.17$.}
	\label{fig_2d_btw_usf_area} 
\end{figure}

As the probability distribution is a homogeneous function of its arguments, we can re-express Eq.~(\ref{eq_prob_1}) as
\begin{subequations} 
	\begin{align}
P(x, x_c) = \frac{1}{x^{\tau_x+\theta}} G(u) = \frac{1}{x_{c}^{\tau_x+\theta}}H(u),
\label{prob_2}
\end{align}
where $u = x/x_c$. In the regime $u\ll 1$, the scaling functions behave as
\begin{align} 
G(u) \sim  u^{\theta} ~~{\rm and}~~  H(u) \sim u^{-\tau_x}.
\end{align}
\end{subequations}
The scaling functions isolate the two exponents. 
As a first example, consider one-dimensional (1d) BTW sandpile model~\cite{Christensen_2005}. Here, $x = s=a$. Numerically, one gets $ P(x,x_c) = 1/x_c$ for $0\le x\le x_c$,
where $x_c=L$.
In the regime $0\leq u \leq 1$, the scaling functions are 
\begin{equation}
G(u) = u~~~~{\rm~~and}~~~~H(u) = 1.
\end{equation} 
Thus, $\theta = 1, \tau_x = 0$, and $D_x =1$. Unlike the 1d percolation model, the 1d BTW model shows trivial behavior, but the FSS reveals a precise scaling of size distribution.

Our proposal is quite simple and well applicable to 1d BTW sandpile. However, this does not seem correctly incorporated in other non-trivial SOC systems. The reason may be partly because of the overwhelming success of the FSS with $\theta = 0$ for a wide range of processes. Also, this feature is not visible in the large-scale system since $\theta$ is typically small. We emphasize that many cases (shown below) can have $\theta \ne 0$. As shown in Fig.~\ref{fig_2d_btw_usf_area}, the scaling functions confirm the existence of system size scaling behavior ($\theta \approx 0.17$) for area distribution in the 2d BTW sandpile model.

In simulations, we collect the avalanches after discarding transients. In models with continuous state variables (energy), we choose the initial configuration close to the critical energy value. Such a choice is helpful in the sense that it can reduce transients. To observe a precise dependence of system size, we keep the total number avalanches $\mathcal{M}$ fixed. As a result, the normalization does not influence PDF. We prefer not to normalize the PDF. It allows us to see, at a fixed $\mathcal{M}$, how large the system size is to consider getting a sharp cutoff. Notice that a clean cutoff is essentially required to determine the cutoff exponent $D_x$. We use log-bin for a relatively large system to avoid losing information near the cutoff.
Thus, $\mathcal{M}$ should be large enough to detect a clean cutoff. To compute the scaling functions, the exponents $\tau_x$ and $\theta$ need to be determined. Alternatively, it is easy to measure $\tau_x +\theta$ by looking at the slope of PDF on a double log scale, with different system sizes, near a larger value of $x$.

\section{Train model}\label{sec_4} 
In the second example, we consider a train model~\cite{Vieira_2000}. The model explains the stick-slip phenomenon and is a simplified version of the spring block model introduced by Burridge and Knopoff~\cite{Burridge_1967}. Interestingly, a recent study~\cite{Naveen_2021} suggests that the train model does not belong to the universality class of the Oslo sandpile model. However, both model show the same critical exponents describing avalanche size distribution~\cite{Paczuski_1996}. We focus on the train model. The model definition is easy to follow. Consider a one-dimensional lattice of size $N$. Assign a continuous force or stress variable $f_i$ to each site $i$ such that $0\le f_i < f_0$, where $f_0$ is a threshold force. We drive the system at one boundary $f_1 \to f_0 + \delta$, with $\delta \ll f_0$. If a site $i$ is unstable $f_i\ge f_0$, it relaxes by transferring a part of the force to neighbors via the following rules: 
\begin{eqnarray}
f_i \to f'_i, \nonumber \\
f_{i\pm 1} \to f_{i\pm 1} + \Delta/2, \nonumber 
\end{eqnarray}
where $\Delta = f_i-f'_i$. In turn, the neighbor(s) may become unstable. It may further trigger the linked sites. This activity forms an avalanche event. The open boundary allows dissipation, and a new avalanche starts when the previous one is completely over.

\begin{figure}[t]
	\centering
	\scalebox{0.69}{\includegraphics{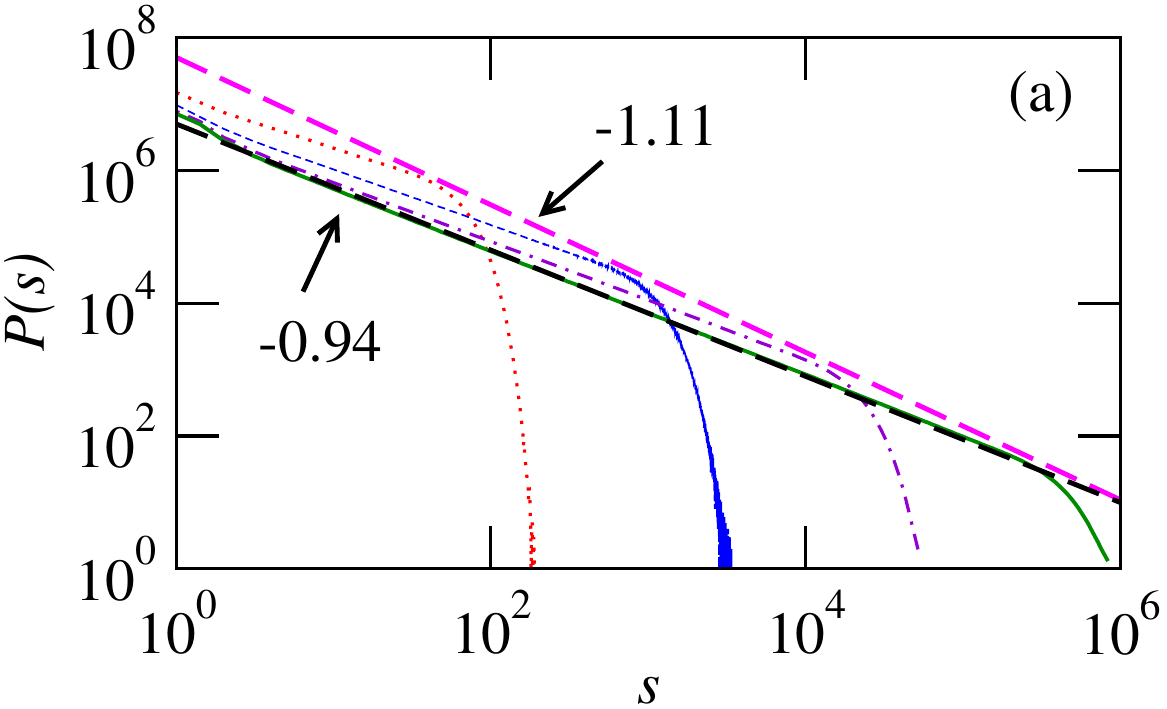}}
	\scalebox{0.69}{\includegraphics{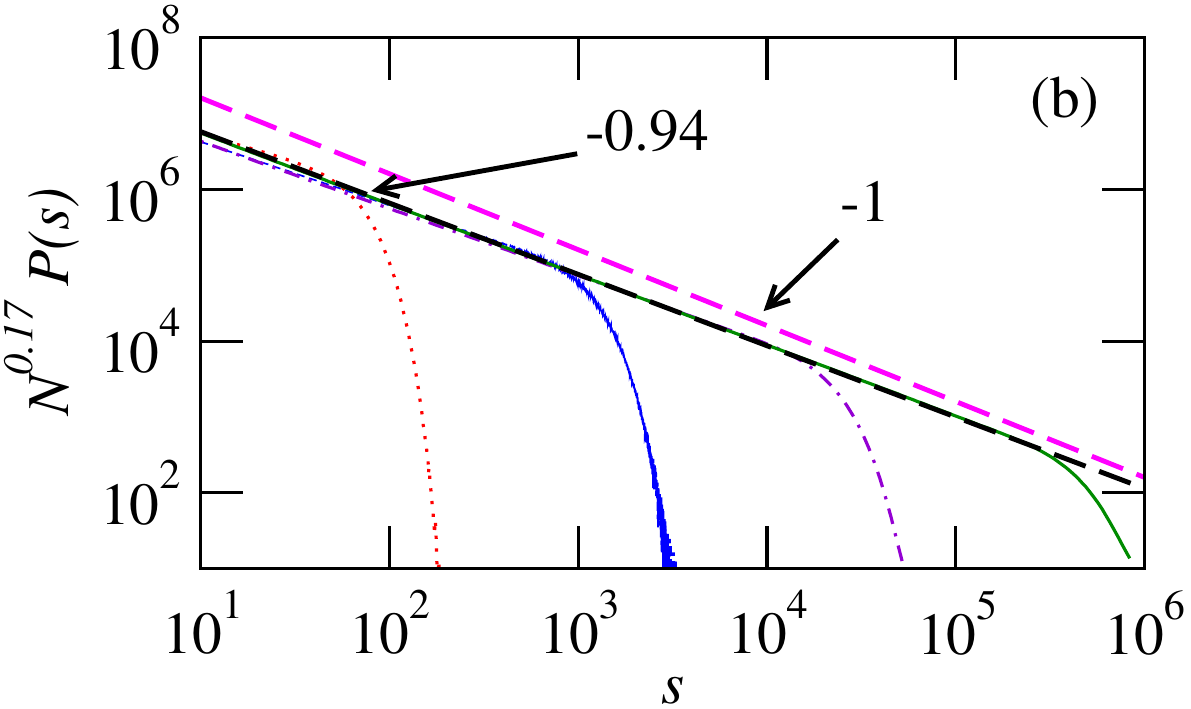}}
	\caption{Bulk driven train model: The avalanche size distribution (a) $P(s)$ (b) $N^{0.17}P(s)$ for different $N$ values $ 2^4, 2^6, 2^8$, and  $2^{10}$. }
	\label{fig_tm} 
\end{figure}

In the deterministic version of the model, $f'_i$ is a nonlinear periodic function. We consider $f'_i$ as a uniformly distributed random variable $\in (0, 1)$. This choice does not change the system properties. We take $f_0 = 1$ and $\delta = 0.1$. 
In the bulk-driven version of the train model, an activation occurs by selecting a random site. 
The numerical results shown in Fig.~\ref{fig_tm} suggest that the avalanche size distribution shows a scaling of $N^{-\theta'}$ type with $\theta' \approx 0.17$. We also numerically checked $\ln(N) P(s)$, but it does not show a good collapse. Also, the scaling exponent $\tau_s \approx 0.94$ is slightly less than 1 and significantly different from 1.11 [as obtained using moment analysis for the Oslo sandpile with bulk drive~\cite{Christensen_2005} or as indicated in Fig~\ref{fig_tm} (a)]. From Eq.~(\ref{eq_prob_1}), it is easy to note that $\theta' = \theta D_s$, where $D_s$ is the cutoff size exponent. The known value of the cutoff size exponent $D_s \approx 2.25$ yields $\theta \approx 0.07$. It turns out that $\tau_s + \theta = 1.01$ is in good agreement with the inequality [cf. Eq.~(\ref{ineq_1})] within the statistical error.

\section{Summary and Discussion}\label{sec_5}
In summary, we have shown that several SOC models, including 2d BTW sandpile, can show explicit system size dependence besides the cutoff in the probability distribution function associated with critical avalanches. Mathematically, the normalization of power-law PDF with an upper cutoff reveals the PDF can be an explicitly power-law or logarithmic function of system size, and the critical avalanche size exponent $\tau_s$ maybe even less than or equal to 1, respectively. Effectively, a simple approximation for the explicit system size function can be a decaying power-law with a scaling exponent $\theta$. The scaling method provides a systematic approach, capturing the finite-size scaling. The scaling functions isolate the critical exponents $\theta$ and $\tau_s$.

We suggest that for 2d BTW sandpile, the exact value of the exponent should be $\tau_s = 1$, as the logarithmic system size dependence arises for avalanche size (area) distribution. Thus, we provide significant insight into one of the intriguing issues associated with SOC. Similarly, we note $N^{-\theta'}$ with $\theta' \approx 0.17$ dependence for the bulk driven train model, and the avalanche size exponent is slightly less than 1. Eventually, the results would be helpful to validate a theoretical prediction and recognize the universality class.
Treating the critical exponent as a function of system size, as suggested in some earlier studies, does not seem convincing.

We have also examined boundary driven train model~\cite{Vieira_2000}, neuronal level model~\cite{Das_2019}, and number-theoretic division model~\cite{Luque_2008}. All these manifest critical avalanches, where $\tau_s$ is significantly greater than 1. In these, we find no evidence of the system size scaling (i.e.,~$\theta=0$).

In percolation, it is the fisher exponent $\tau'_{s} =\tau_s+1$ that conventionally takes a value greater than 2. Several recent works~\cite{Alvarado_2013, Sheinman_2015, Pruessner_2016, Sheinman_2016, Ziff_2016, Christensen_2008} have shown that the exponent can take a value less than 2 in many physically interesting systems. Examples include no-enclave percolation describing the behavior of active gel driven internally by molecular motors~\cite{Alvarado_2013} and percolation on not visited sites for a 2d random walk~\cite{Amit_2021}. Also, the exponent is $\tau'_{s}<2$ in the forest-fire model~\cite{Schenk_2002}. While a clean detection of logarithmic corrections for $\tau_s = 1$ may be difficult, it is easy to verify the explicit system size dependence numerically.

\section*{ACKNOWLEDGMENTS}
ACY acknowledges seed grants under IOE and SERB, DST, Government of India (Grant No. ECR/2017/001702) for their support.
AQ acknowledges the Department of Science and Technology, Government of India for Inspire Fellowship (DST/INSPIRE Fellowship/IF180689). HHJ acknowledges support from University Grants Commission (UGC), India.

\end{document}